\RequirePackage{fix-cm}
\documentclass{svjour3}                     
\smartqed  
\usepackage{graphicx}
\usepackage{bm}
\usepackage{amsmath}	
\usepackage{braket}

\begin{document}

\title{The isoscalar monopole strength of $^{13}{\rm C}$} 
\author{S. Shin\and B. Zhou \and M. Kimura}

\institute{S. Shin \at
           Department of Physics, Hokkaido University, Sapporo 060-0810, Japan. \\
           \email{shin@nucl.sci.hokudai.ac.jp}
           \and
           B. Zhou \at
	   Institute of Modern Physics, Fudan University, Shanghai 200433, China. \\
	   \email{zhou\_bo@fudan.edu.cn} 
	   \and
           M. Kimura \at
           Department of Physics, Hokkaido University, Sapporo 060-0810, Japan. \\
	   Nuclear Reaction Data Centre, Hokkaido University, Sapporo 060-0810, Japan.\\
	   Research Center for Nuclear Physics (RCNP), Osaka University, Ibaraki 567-0047, Japan\\
	   \email{masaaki@nucl.sci.hokudai.ac.jp} 
           }
\date{Received: date / Accepted: date}

\maketitle

\begin{abstract}
 To identify the 3$\alpha$ BEC state with the excess neutron, we have investigated the monopole
 strength of the excited states of $^{13}{\rm C}$ by using the theoretical framework of the
 real-time evolution method. The calculations have revealed several candidates of the Hoyle-analog
 states in highly excited region.  
\end{abstract}

\section{Introduction}\label{sec:intro}
In this decade, it has been intensively discussed that the Hoyle state (the $0^+_2$ state of  
$^{12}{\rm C}$) is a dilute gas-like $\alpha$ cluster
state~\cite{Uegaki1978,Kamimura1981,Kanada-Enyo1998,Chernykh2007} and  regarded as a BEC of three $\alpha$ 
particles~\cite{Tohsaki2001,Schuck2016}.  The idea of the $\alpha$ particle condensate has been
extended to the other  excited states of $^{12}{\rm C}$. Namely, the $2^+$ state at 10.03
MeV~\cite{Freer2009,Itoh2011} and  the $4^+$ state at 13.3 MeV~\cite{Freer2011} are expected as the
rotational excitation mode of the  Hoyle state and assigned as the "Hoyle
band"~\cite{Freer2014a,Funaki2015}. More recently, the $0^+_3$ state at 10.3  MeV has been suggested
as  the "breathing mode" of the Hoyle state~\cite{Kurokawa2005,Ohtsubo2013,Funaki2015,Zhou2016}. 

The Hoyle-analog states in neighboring nuclei have also been the important subjects. In 
particular, $^{13}{\rm C}$ is an interesting system as it is composed of three $\alpha$ particles
(bosons) and an excess neutron (fermion). It is a fascinating question how the 3$\alpha$ BEC is
disturbed by the addition of a neutron as an impurity. In order to answer this question, several
experimental and theoretical studies have been
conducted~\cite{Furutachi2011,Wheldon2012,Yamada2015,Ebran2017,Chiba2020}, but
the conclusion still remains controversial.      

Recently a theoretical model named the real-time evolution method (REM)~\cite{Imai2019,Zhou2020} has
been proposed as a powerful approach to the $\alpha$ clustering of light nuclei. It was demonstrated
that the model precisely describes the Hoyle state (3$\alpha$ BEC). Furthermore, the model has also been
applied to the low-lying states $^{13}{\rm C}$ to investigate the underlying symmetry of the nuclear
shape~\cite{Shin2021}. Therefore, it is natural to extend the discussion to the highly excited states of
$^{13}{\rm C}$ to search for the Hoyle-analog state (3$\alpha$ BEC plus an excess neutron). In this
work, we report the structure of the highly excited states of $^{13}{\rm C}$ calculated by
REM. We focus on the excited $1/2^-$ states and their monopole strengths as a signature of
the BEC formation.   

\section{Theoretical Framework}\label{sec:framework}
The Hamiltonian and the theoretical framework used in this work are essentially same with those in
Ref.~\cite{Shin2021}. The Hamiltonian is given as, 
\begin{align}
 \hat{H} = \sum_{i=1}^{13} \hat{t}_i  + \sum_{i<j}^{13}\hat{v}_N(r_{ij}) +
 \sum_{i<j}^{13}\hat{v}_C(r_{ij}) - \hat{t}_{cm},  
 \label{eq:ham}
\end{align}
where $\hat{t}_i$ and $\hat{t}_{cm}$ are the kinetic energies of nucleon and that of the
center-of-mass, respectively. The nucleon-nucleon interaction $\hat{v}_{N}$ is composed of
the Volkov No.~2 force~\cite{Volkov1965} with the exchange parameters $W=0.4$, $B=H=0.125$ and
$M=0.6$, and the spin-orbit part of the G3RS force~\cite{Yamaguchi1979} with the strength of
$u_{ls}=1000$ MeV. 

The basis wave function of REM is the Brink-Bloch wave function~\cite{Brink1966} which consists of
three $\alpha$ clusters with $(0s)^4$ configuration coupled with an excess neutron, 
\begin{align}
 \Phi(\bm Z_1,...,\bm Z_4) &= \mathcal A
 \Set{\Phi_\alpha(\bm Z_1)\Phi_\alpha(\bm Z_2)\Phi_\alpha(\bm Z_3)\phi(\bm r,\bm Z_4)\chi_{n\uparrow}},
 \label{eq:brink1}\\ 
 \Phi_\alpha(\bm Z) &= \mathcal A
 \Set{\phi(\bm r_1,\bm Z)\chi_{p\uparrow}\cdots\phi(\bm r_4,\bm Z)\chi_{n\downarrow}},\\
 \phi(\bm r,\bm Z) &= \left({2\nu}/{\pi}\right)^{3/4}\exp
 \set{-\nu\left(\bm r- \bm Z\right)^2},
\end{align}
where $\Phi_\alpha(\bm Z_1)...\Phi_\alpha(\bm Z_3)$ describe the $\alpha$ clusters. The excess
neutron is located at $\bm Z_4$ and its spin is fixed to up. The size parameter $\nu$  is fixed to
0.235 $\rm fm^{-2}$ to reproduce the radius of $\alpha$ particle. We calculate the real-time
evolution of $\bm Z_1,...,\bm Z_4$ by solving the equation-of-motion, and obtain a set of the basis
wave functions. 
\begin{align}
  i\hbar&\sum_{j,\sigma} C_{i\rho j\sigma}\frac{dZ_{j\sigma}}{dt} =
 \frac{\partial \mathcal H}{\partial Z_{i\rho}^*}, \quad
 \mathcal H\equiv\frac{\braket{\Phi|\hat{H}|\Phi}}{\braket{\Phi|\Phi}},\quad
 C_{i\rho j\sigma}\equiv\frac{\partial^2}{\partial Z^*_{i\rho}\partial Z_{j\sigma}}
 \ln\braket{\Phi|\Phi}.
\end{align}
Then, we apply the generator coordinate method (GCM) superposing the wave functions,
\begin{align}
 \Psi^{J\pi}_{M}= \sum_{iK}\hat{P}^{J\pi}_{MK}f_{iK}
 \Phi(\bm Z_1(t_i),...,\bm Z_4(t_i)), \label{eq:gcmwf2}
\end{align}
where $\hat{P}^{J\pi}_{MK}$ is the parity and the angular momentum projector and the real-time $t$
is discretized. The amplitude $f_{iK}$ and eigen-energy are determined by solving the Hill-Wheeler
equation~\cite{Hill1953}. 

From thus-obtained wave functions, we calculate the monopole transition matrix elements as a
signature of the Hoyle-analog state,
\begin{align}
 M_{E0, IS0} = 
 \braket{\Psi^{1/2^-}_{M}({\rm ex.})|\mathcal M_{E0,IS0}|\Psi^{1/2^-}_{M}({\rm g.s.})}, 
\end{align}
where $\Psi^{1/2^-}_{M}({\rm g.s.})$ and $\Psi^{1/2^-}_{M}({\rm ex.})$ denote the wave functions of
the ground state and excited $1/2^-$ states, respectively. $\mathcal M_{E0,IS0}$ is either of the
electric (E0) or isoscalar (IS0) monopole operator,
\begin{align}
 \mathcal{M}_{E0} = e^2\sum_{i=1}^A r_i^2\frac{1+\tau_z^{(i)}}{2}, \quad 
 \mathcal{M}_{IS0} = \sum_{i=1}^A r_i^2.
\end{align}

\section{Results and discussions}\label{sec:results}
\begin{figure}[hbt]
 \includegraphics[width=0.9\hsize]{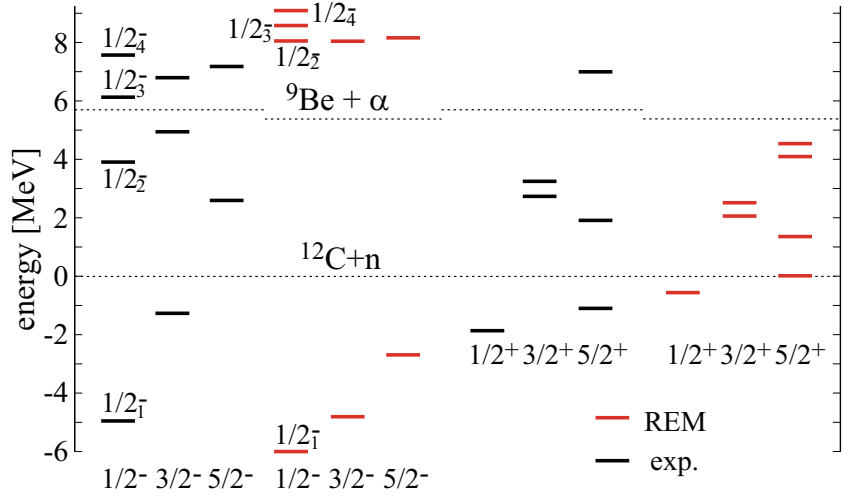}
\caption{Calculated and observed partial level scheme of $^{13}{\rm C}$. The energy is measured
 relative to the $^{12}{\rm C}+n$ threshold.}
\label{fig:1}       
\end{figure}

Figure~\ref{fig:1} shows the calculated and observed spectra of the $1/2^\pm$, $3/2^\pm$ and
$5/2^\pm$ states, which is essentially same with that discussed in our previous work~\cite{Shin2021}. 

Although the calculation yields the correct order of the ground band  ($1/2^-$, $3/2^-$ and $5/2^-$
states), it overestimates the moment-of-inertia. Furthermore, it does not yields some of the
observed negative-parity states lying between the $^{12}{\rm C}+n$ and $^{9}{\rm Be}+\alpha$
thresholds. These discrepancies are due to the inability of the model in describing the distortion
the cluster structure as we have assumed 3$\alpha+n$ cluster structure. 
On the contrary, the calculation describes the positive-parity spectrum relatively better than the
negative-parity one. This indicates that the positive parity states have more pronounced 3$\alpha+n$
cluster structure than the low-lying negative-parity states. 

Now we turn to the highly excited states, in particular, the $1/2^-$ states. They can have enhanced
monopole transition strength which is regarded as a signature of the enhanced clustering. Indeed, an
experiment has been conducted to measure the monopole transition strength and it reported three
$1/2^-$ states as candidates of the pronounced 3$\alpha+n$ clustering. 
Table~\ref{tab:1} summarizes the properties of the ground state and the excited $1/2^-$ states. 
\begin{table}[hbt]
\caption{The properties of the ground and excited $1/2^-$ states; excitation energies; matter,
 neutron and proton root-mean-square radii;  isoscalar and electric monopole transition
 matrices. The energies, radii, isoscalar and electric transition matrices   are given in the unit
 of MeV, fm, $\rm fm^2$ and $e\rm fm^2$, respectively. The observed data~\cite{Kawabata2008} are
 also given.}  
\label{tab:1}       
\begin{tabular}{c|cccccc|cc}
\hline
 &\multicolumn{6}{c}{REM} & \multicolumn{2}{|c}{exp.}\\
\hline
 & $E_x$  & $\sqrt{\braket{r^2_m}}$ & $\sqrt{\braket{r^2_n}}$ & $\sqrt{\braket{r^2_p}}$ 
 & $M_{IS0}$   & $M_{E0}$ & $E_x$ & $M_{IS0}$\\
\hline
 g.s. & 0.0  & 2.44 & 2.49 & 2.39 & --    & --   & 0.0   & --\\
 ex.  & 14.1 & 3.14 & 3.56 & 2.57 & 6.0   & 2.2  & 8.86  & 6.1\\
      & 14.6 & 3.03 & 3.34 & 2.61 & 3.7   & 1.1  & 11.08 & 4.2\\
      & 15.1 & 3.00 & 3.27 & 2.63 & 12.5  & 5.6  & 12.50   & 4.9\\
\hline\end{tabular}
\end{table}
It is notable that all of the calculated $1/2^-$ states have considerably enhanced monopole strengths
comparable with the Weisskopf estimate (1WU = 5.9 $e\rm fm^2$), and show reasonable agreement
with the data reported by an experiment~\cite{Kawabata2008}. Because of the enhanced monopole strength, they
are regarded as the candidate of the Hoyle-analog states. In fact, their matter radii are
much larger than the ground state  showing their dilute density. 
However, the proton radii are not as large as that of the Hoyle state which is calculated as 3.7
fm by using the same Hamiltonian~\cite{Kamimura1981}. This implies the reduction of the  3$\alpha$
BEC size due to the attraction between the $\alpha$ cluster and the excess neutron. It is also noted
that the present calculation overestimates the observed excitation energies of these states, that
tells us  a need for improvement in our Hamiltonian and model wave functions.  

To elucidate the internal structure of the ground and excited $1/2^-$ states, we have calculated 
the overlap between the $1/2^-$ states and Brink-Bloch wave functions, which is defined as,
\begin{align}
 |\braket{\Psi^{1/2^-}_{n}|P^{1/2^-}_{MK}\Phi(\bm Z_1(t_i),...,\bm Z_4(t_i))}|^2,
\end{align}
where $\Psi^{1/2^-}_n$ is the wave function of the $1/2^-_{1-4}$ states, while 
$\Phi(\bm Z_1(t_i),...,\bm Z_4(t_i))$ is the basis wave function.
%
\begin{figure}[hbt]
 \includegraphics[width=\hsize]{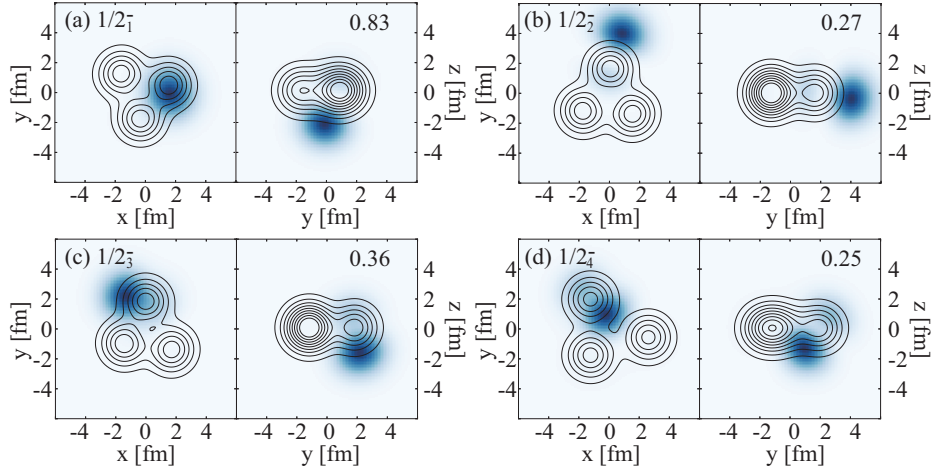}
 \caption{Density distributions of the Brink-Bloch wave functions which have maximum overlap with 
 the $1/2^-_{1-4}$ states. Solid lines and color plots show the density distributions of 3$\alpha$
 particles and the excess neutron, respectively. Each panel show the densities in $xy$ and $yz$
 planes. Numbers in the panels show the maximum value of the overlap.}
\label{fig:2}       
\end{figure}
The ground state maximally overlaps with the Brink-Bloch wave function shown in Fig.~\ref{fig:1}
(a). Note that the overlap is as large as 0.83, and hence, the ground state may be reasonably
approximated by this wave function which has relatively compact spatial distribution. On the
contrary, the excited $1/2^-$ states only have the maximum overlaps ranging  0.25$\sim$0.36 with the
wave functions in panels (b)-(d). This clearly indicates that the excited states cannot be
approximated by a single Brink-Bloch wave function as they have non-localized cluster
configurations. Indeed,  these excited states also overlap with various Brink-Bloch wave functions
with different cluster configurations indicating that they do not have definite nuclear shape nor
underlying spatial symmetry. This feature is common with the Hoyle state and reasonably in
accordance with the dilute gas-like nature. Thus, within the scope of our study, these three excited
states are good candidates of the Hoyle-analog state. 

The main difference from $^{12}{\rm C}$ is that there are three BEC candidates of $^{13}{\rm C}$ in
both the experimental and theoretical results.  To identify the BEC state precisely, we need to
investigate the spectroscopic factors and occupation probabilities, and such calculation is ongoing.

\section{Summary}\label{sec:summary}
We have investigated the highly excited states of $^{13}{\rm C}$ and their monopole transition
strengths to identify the Hoyle-analog state. We have obtained three candidates which have enhanced
monopole transition strengths from the ground state. The calculated monopole strength are as large
as those reported by an experiment, although the calculation overestimated the excitation energies. 
The calculation also showed that the radii of these candidates are considerably smaller than that
of the Hoyle state. This implies that the extra neutron causes the shrinkage of 3$\alpha$ BEC due to
its interaction with $\alpha$ particles.

\begin{acknowledgements}
 The authors acknowledge the fruitful discussions with Dr. Funaki and  Dr. Kawabata. This work was
 supported by JSPS KAKENHI Grant Nos. 19K03859, the collaborative research programs 2021 at the
 Hokkaido University information initiative center. The calculations have been performed by the
 computers  in YITP at Kyoto University and RCNP at Osaka University.
\end{acknowledgements}

\bibliographystyle{spphys}       
\bibliography{C13.mono}   


\end{document}